\documentstyle[psfig]{mn}

\title[Optical and Near-infrared Imaging of faint GPS Sources]{Optical and near-infrared imaging of faint \\ Gigahertz Peaked Spectrum sources}

\author[I. Snellen et al.]{I.A.G. Snellen$^{1,2}$, R.T. Schilizzi$^{1,3}$,
M.N. Bremer$^{1,6}$,A.G. de Bruyn$^{4,5}$, G.K. Miley$^{1}$, \cr
H.J.A. R\"ottgering$^{1}$, R.G. McMahon$^{2}$, I. P\'erez Fournon$^{7}$\\ 
$^{1}$Leiden Observatory, P.O. Box 9513, 2300 RA, Leiden, The Netherlands \\
$^{2}$Institute of Astronomy, Madingley Road, Cambridge CB3 0HA, United Kingdom\\
$^{3}$Joint Institute for VLBI in Europe, Postbus 2, 7990 AA, Dwingeloo, The Netherlands\\
$^{4}$Netherlands Foundation for Research in Astronomy, Postbus 2, 7990 AA, 
Dwingeloo, The Netherlands\\
$^{5}$Kapteyn Institute, Postbus 800, 9700 AV, Groningen, The Netherlands\\
$^{6}$Institut d'Astrophysique de Paris, 98bis Boulevard Arago, 75014 
         Paris, France \\
$^{7}$Instituto de Astrofisica de Canarias, Tenerife, Spain}

\date{}
\begin{document}
\maketitle

\begin{abstract}
A sample of 47 faint Gigahertz Peaked Spectrum (GPS) radio sources 
selected from the Westerbork Northern Sky Survey (WENSS, Rengelink et al. 
1997), has been imaged in the optical and near infrared, 
resulting in an identification fraction of 
87\%. The $R$$-$$I$ and $R$$-$$K$ colours of the faint optical counterparts 
are as expected for passively evolving elliptical galaxies, assuming that they
 follow the $R$ band
Hubble diagram as determined for radio-bright GPS galaxies.
We have found evidence that the radio spectral properties of the
GPS quasars are different from those of GPS galaxies:
The observed distribution of radio spectral peak frequencies  for
 GPS sources optically identified with 
bright stellar objects (presumably quasars) is shifted compared with
 GPS sources  identified with faint or extended optical objects
(presumably galaxies), in the sense that a GPS quasar is likely to 
have a higher peak frequency than a GPS galaxy. 
This means that the true peak frequency distribution 
is different for the GPS galaxies and quasars, because the sample 
selection effects are independent of optical identification. 
The correlation between peak frequency and redshift as has been suggested for
bright sources
has not been found in this sample; no correlation exists between $R$ magnitude 
(and therefore redshift) and peak frequency for the GPS galaxies.
We therefore believe that the claimed correlation is actually caused by the
 dependence of the peak frequency on optical host, because the GPS
galaxies are in general at lower redshifts than the quasars.
The difference in the peak frequency distributions of the GPS galaxies and 
quasars is further evidence against the hypothesis that they form a 
single class of object.
\end{abstract}

\section{Introduction}
Extragalactic radio sources have been studied for many years, but it is still
unclear how they are formed and evolve. A crucial element in the study of
their evolution is to identify the young counterparts of the ``old'' FRI/FRII
extended objects. Good candidates for
young radio sources are those with peaked spectra (Gigahertz Peaked Spectrum
(GPS) sources and Compact Steep Spectrum (CSS) sources,
Fanti et al. 1995, Readhead et al. 1996, O'Dea and Baum 1997, O'Dea 1998), 
because they are
small in angular size as expected for young sources. GPS sources are
characterized by a convex radio spectrum peaking at a frequency of about 1 GHz
and are typically 100 pc in size. CSS sources have a peak in their spectrum at
lower frequencies and have projected linear sizes of $< 10 - 15$ Kpc.
A commonly discussed alternative to them being young, is that they are
small due to confinement by a 
particularly dense and clumpy interstellar medium that impedes the 
outward propagation of the jets (O'Dea et al 1991). 
 
Important information about the nature of Gigahertz Peaked Spectrum GPS
sources comes from the properties of their optical counterparts.
Systematic optical identification programs on GPS sources have
been performed by O'Dea et al. (1990), Stanghellini et al. (1993) and de Vries
et al. (1995), and these show that the optical counterparts  are a mixture of
galaxies and quasars. O'Dea et al. (1990) and Stanghellini et al. (1993) 
found that disturbed optical morphologies are a common characteristic of GPS
galaxies, with many of them having close companions or appearing to lie in a
group or cluster. This may indicate that galaxy-galaxy interactions and/or
mergers play an important role in the GPS phenomenon.
GPS galaxies appear to be a homogeneous class of giant ellipticals with old 
stellar populations (Snellen et al. 1996a, 1996b, O'Dea et al. 1996). 
Their optical to near-infrared fluxes seem not to be contaminated by light
produced by the active nucleus, and they are therefore potentially useful 
probes of galaxy evolution. 

In the light of the orientation unification scheme (eg. Barthel 1989), it 
would be expected that the redshift distributions of GPS sources identified 
with galaxies and quasars are more or less the same.
However, the GPS quasars tend to be at much higher redshift than
the galaxies ($2<z<4$, O'Dea et al. 1991). In addition, the intrinsic 
peak frequency distributions and radio morphologies seem to be different
for GPS galaxies and quasars (Stanghellini et al. 1996). 
These results are probably influenced by several selection effects, but
if true, they may indicate that GPS galaxies and quasars are not unified by
orientation, but form separate classes of object which happen to have similar
radio spectra.

To further address these issues, we have started to investigate GPS
sources at fainter flux density levels (Snellen et al. 1995, Snellen et 
al. 1998) than those in the samples of bright GPS sources
studied until now (O'Dea et al. 1991, de Vries et al. 1997, Stanghellini et 
al. 1998).
By comparing the properties of these new faint samples with those of bright 
samples in the literature, we shall investigate properties of GPS sources as a
function of radio luminosity, redshift and rest frame peak frequency.

This paper describes the optical and near-infrared imaging of a sample 
of faint GPS sources selected from the Westerbork Northern Sky Survey 
(WENSS Rengelink et al. 1997).

\section{The Sample of Faint GPS Sources}

The selection of the sample has been described in  detail in Snellen et al. 
1998 (paper I).
Candidate GPS sources selected from the Westerbork Northern Sky survey, are
those with an inverted spectrum between 325 MHz and higher frequencies.
The
sources are located in two regions of the survey; one at $15^h < \alpha < 
20^h$
and $58^\circ< \delta < 75^\circ$, which is called the {\it mini-survey} region
(Rengelink et al. 1997), and the other at $4^h00^m < \alpha < 8^h30^m$ and
$58^\circ< \delta < 75^\circ$. Additional observations at 1.4, 5, 8.4 and 15
GHz were carried out with the Westerbork Synthesis Radio Telescope (WSRT) and 
the Very Large Array (VLA), yielding a sample of 47
genuine GPS sources with peak frequencies ranging from 500 MHz to more than 15
GHz, and peak flux densities ranging from $\sim30$ to $\sim900$ mJy.
To determine whether the GPS sources have an optical counterpart on the 
Palomar Sky Survey plates (POSS), the Automatic Plate Measuring (APM) 
catalogue (Irwin et al. 1994) and the HST Guide Star Catalogue 
(Russel et al. 1990) were used for the sources in the mini-survey region and
in the region $4^h00^m < R.A. < 8^h30^m$, respectively.
This information was used for finding charts, to determine 
the required exposure times for the observations, and for astrometric purposes.

\section{Observations}

The sample was imaged in the optical in the course of three 
observing sessions
(see table \ref{observations}), with the 0.9m 
Jacobus Kapteyn Telescope (JKT),
the 2.5m Isaac Newton Telescope (INT) and the 2.5m Nordic Optical Telescope
(NOT), all located at Observatorio del Roque de los Muchachos on La Palma, 
Spain. The sample was also observed in the near-infrared with 
the 4.2m William Herschel Telescope
(WHT), and with the NOT.
The INT, NOT and WHT observations were done as part of the international CCI
observing time awarded to the WENSS team in 1995.

During the different observing sessions, several standard stars from 
Landolt (1983) were observed for photometric calibration.
A summary of the relevant parameters for the observations is given in table
\ref{observations}.

\begin{table*}

\begin{tabular}{|rrrlcrc|}\hline
Telescope&Observing&Detector&Filter& Detector&Pixel & Field of\\
         &Date     &&&  Size    &Size          & View    \\ \hline
JKT &3-9 Jun. 1994   &CCD&R, I   &$400\times590$&    $0.30''$&$2'\times 3'$\\
INT &  6 Apr. 1995   &CCD&R      &$1024\times1024$&   $0.58''$&$10'\times10'$\\
NOT &1-5 Oct. 1995   &CCD&R      &$1024\times1024$& $0.17''$&$3'\times3'$\\ 
WHT &28\&31 Jan. 1996&WHIRCAM&K  &$256\times256$&$0.24''$&$1'\times1'$\\
NOT & 15-17 Aug. 1995&ARNICA&J, K&$256\times256$&$0.29''$&$1.25'\times1.25'$\\ 
NOT & 5-6 Sep. 1995  &ARNICA&J, K&$256\times256$&$0.29''$&$1.25'\times1.25'$\\ \hline
\end{tabular}
\caption{\label{observations} Details of the optical and near-infrared
observations. }
\end{table*}

\medskip
\noindent {\it JKT CCD Imaging}
\medskip

Six nights of observing time were obtained on the JKT, from 3 to 9 June 1994. All
25 GPS sources in the sample from the mini-survey region were observed in 
$R$ and $I$ bands using Harris filters (Matura et al. 1993), 
which have, in combination with the GEC7
CCD, effective wavelengths of 6550 and 8260 \AA \ 
and FWHM bandwidths of 1500 and 3000 \AA\ for the $R$ and $I$ filters
respectively. Typical observing times were $3\times 300$ seconds for the POSS
detections and $3 \times 600$ seconds for the unidentified sources on the POSS.
The $400\times590$ GEV7-chip used as the detector 
has a projected pixel size
corresponding to 0.3 $''$, and a total field of view of $2'\times3'$. 
In addition to the Landolt standards, spectroscopic standards were measured 
to determine the absolute flux density scale for the particular set of 
filters. 
Limiting surface brightnesses ($1\sigma$) were for the longest exposures 
typically 26.1 and 25.3
mag/arcsec$^{-2}$ in respectively $R$ and $I$ band.
 The weather was
photometric during the run, except for the second half of the night on 7 June.
The sources observed on 7 June were re-observed briefly on 
9 June for photometric calibration purposes.

\medskip
\noindent {\it INT CCD Imaging}
\medskip

Sources which had not been detected with the JKT were reobserved
using the INT on 6 April 1995. These objects, B1525+6801, B1551+6822, B1600+7131, 
B1620+6406, B1639+6711, B1655+6446, and B1808+6813, were observed in $R$ band
using a 1024$\times$1024 TEK-chip and Harris filter 
($\lambda_{eff} = 6550 \AA$, FWHM=1500 \AA).
 Each source was
observed for 15 minutes under photometric conditions.
Limiting brightnesses ($1\sigma$) were typically 27.3 
mag/arcsec$^{-2}$ for the longest exposures.
The CCD detector had a pixel scale of $0.58''$ and a field of view of
$10'\times10'$. 

\medskip
\noindent {\it NOT CCD Imaging}
\medskip

Sources in the region $4^h00^m < R.A. < 8^h30^m$ and
$58^\circ< decl. < 75^\circ$ were observed using the NOT
in the period 1 to 5 October 1995. A 1024$\times$1024 TEK Chip in combination 
with a Harris $R$ band filter ($\lambda_{eff} = 6550 \AA$, FWHM=1500 \AA)
 was used. The CCD detector had a pixel 
scale of $0.17''$ and a field of view of $3'\times3'$.
One source, B1954+6146, was also observed using an $I$
band filter which, in combination with the TEK CCD, had 
an effective wavelength of 8300~\AA.
Limiting brightnesses ($1\sigma$) were typically 27.0 
mag/arcsec$^{-2}$ for the longest exposures.
The conditions were photometric during the whole run.

\medskip
\noindent {\it NOT ARNICA Observations}
\medskip

About half of the  GPS sources in the mini-survey region were observed in $K$
band ($\lambda = 2.20 \pm 0.20 \mu m$), and some in $J$ band 
($\lambda = 1.28 \pm 0.15 \mu m$)using the ARcetri Near Infrared CAmera 
(ARNICA) 
on the NOT, from 15 to 17 August 
and on 5 and 6 September 1995. These observations are discussed in detail
by Villani and di Serego Alighieri (1998). ARNICA utilizes a NICMOS
3 (256 $\times$ 256 pixels) array as a detector, with a projected pixel size 
corresponding to $0.29''$ and a
field of view of $75''\times75''$. In order to ensure background limited
images and to avoid saturation of the brightest sources, the total
exposure time on each source 
was divided into a number of shorter exposures,
between which the telescope was moved in a 9 position raster.
The total exposure times were between 6 and 30 minutes, depending on the 
expected magnitude of the object.

\medskip
\noindent {\it WHIRCAM Observations}
\medskip

Seventeen sources in the region $4^h00^m < \alpha < 8^h30^m$ were observed on 28
and 31 January 1996, using the WHT Infrared Camera (WHIRCAM). WHIRCAM utilizes
a $256\times256$ InSb array detector, and was used as a direct camera with an
image scale of 0.24 arcsec/pixel. Each source was observed in $Kshort$ band
($\lambda = 2.16 \pm 0.16 \mu m$) with an 8 position jitter pattern. At each
position the field was integrated for 2, 4 or $10 \times 10$ seconds using the
ND\_STARE mode, meaning that the array is reset and read immediately and
read again after the exposure time. This resulted in total exposure times of
160, 320 or 800 seconds, depending on the expected magnitude of the object.
Two sources, B0830+5813 and B0755+6354, were observed on 28 January 1996 
through thin cirrus. The conditions on 31 January were photometric.

\subsection{Optical Data Reduction and Calibration}

The reduction of the optical observations 
was carried out with the software packages IRAF from the 
NOAO and AIPS from the NRAO. The fits images were loaded into IRAF and 
bias subtracted and flat fielded in the standard manner. 
Positional calibration was then carried out within AIPS using the task
XTRAN. Positions of stars in the CCD fields were taken from the APM 
catalogue (Irwin et al. 1994) for the sources in the mini-survey
region and the HST Guide Star Catalogue for the sources with 
$4^h00^m < \alpha < 8^h30^m$. These positions have an intrinsic rms accuracy 
of typically $0.7''$. The epoch and equinox 
of the reference coordinate system was then
converted from B1950 to J2000 using the AIPS task EPOSWITCH.

The absolute flux density scale was determined using the spectroscopic
standard stars observed during the JKT run. This was done by combining the
tabulated stellar spectra (Stone 1977) with the filter relative transmission 
curves and the CCD
response. An object of zero magnitude, observed in the  Harris $R$ band
filter, was estimated to have a flux density of $2.2\pm0.2\times10^{-9}$ $erg$
$cm^{-2}$ $sec^{-1}$ $\AA ^{-1}$ at 6550 \AA . An object of zero magnitude,
observed in the  Harris $I$ band filter,  has been estimated to have a flux
density of $1.0\pm0.1\times10^{-9}$ $erg$ $cm^{-2}$ $sec^{-1}$ $\AA ^{-1}$ 
at 8900 \AA . The filter response curves indicate that these absolute
calibration measurements can also be applied to the Harris $R$ band
observations on the NOT and the INT, due to the similar CCD characteristics. 
The $I$ band filter used on the
NOT had no absolute photometric calibration information available. However,
since the filter bandpass is similar to the Harris I band filter used on the
JKT, the same absolute calibration was used.

\subsection{Near-infrared Data Reduction and Calibration} 

The data of the ARNICA and WHIRCAM observations were reduced using IRAF.
The images were dark-subtracted and then flat-fielded by median filtering the
image stack. Thereafter the frames were normalized and mosaiced 
together to produce
one final image per object. In the case of the ARNICA observations, it was
necessary to do the background subtraction from each frame before
flat-fielding, in order to eliminate the contribution of the thermal emission
from the background and the telescope. UKIRT faint standards were observed
several times during the nights at a range of air-masses. Although the WHIRCAM
observations were done using a $K$short filter, the $K$ band magnitude system 
has been
used for which zero magnitude corresponds to a flux density of
$4.1\pm0.4\times10^{-11}$ $erg$ $cm^{-2}$ $sec^{-1}$ $\AA ^{-1}$. 
Astrometric measurements have been
done in the same way as in the optical, using positions of objects in the $R$
band CCD images that were also in the near-infrared fields.

\subsection{Identification Procedure}

The procedure for identifying
a GPS source in the optical is simplified compared to large size radio sources,
because the angular size of the radio source is very small compared
to the optical resolution. The likelihood method (eg. de Ruiter et al. 1977)
was used to quantify our identification procedure.
The reliability of a proposed identification has been assessed using 
the observed differences between the optical and radio
positions and their errors and the optical background counts.
For each candidate counterpart the following dimensionless measure of the 
uncertainty in position difference, $r$, was calculated.

\begin{equation}
r = \sqrt{\frac{\Delta \alpha ^2}{\sigma _{\alpha opt}^2} 
+ \frac{\Delta \delta ^2}{\sigma _{\delta opt} ^2}}
\end{equation}
where $\Delta \alpha$ and $\Delta \delta$ are the offsets between the 
optical and radio positions, and $\sigma _{\alpha opt}$ and 
$\sigma _{\delta opt}$ the uncertainties in the 
optical right ascension and declination positions. 
The error in the radio position is small compared to
the error in optical position ($0.2''$ versus $0.7''$) and was therefore
neglected. 

\begin{figure*}
\hbox{
\psfig{figure=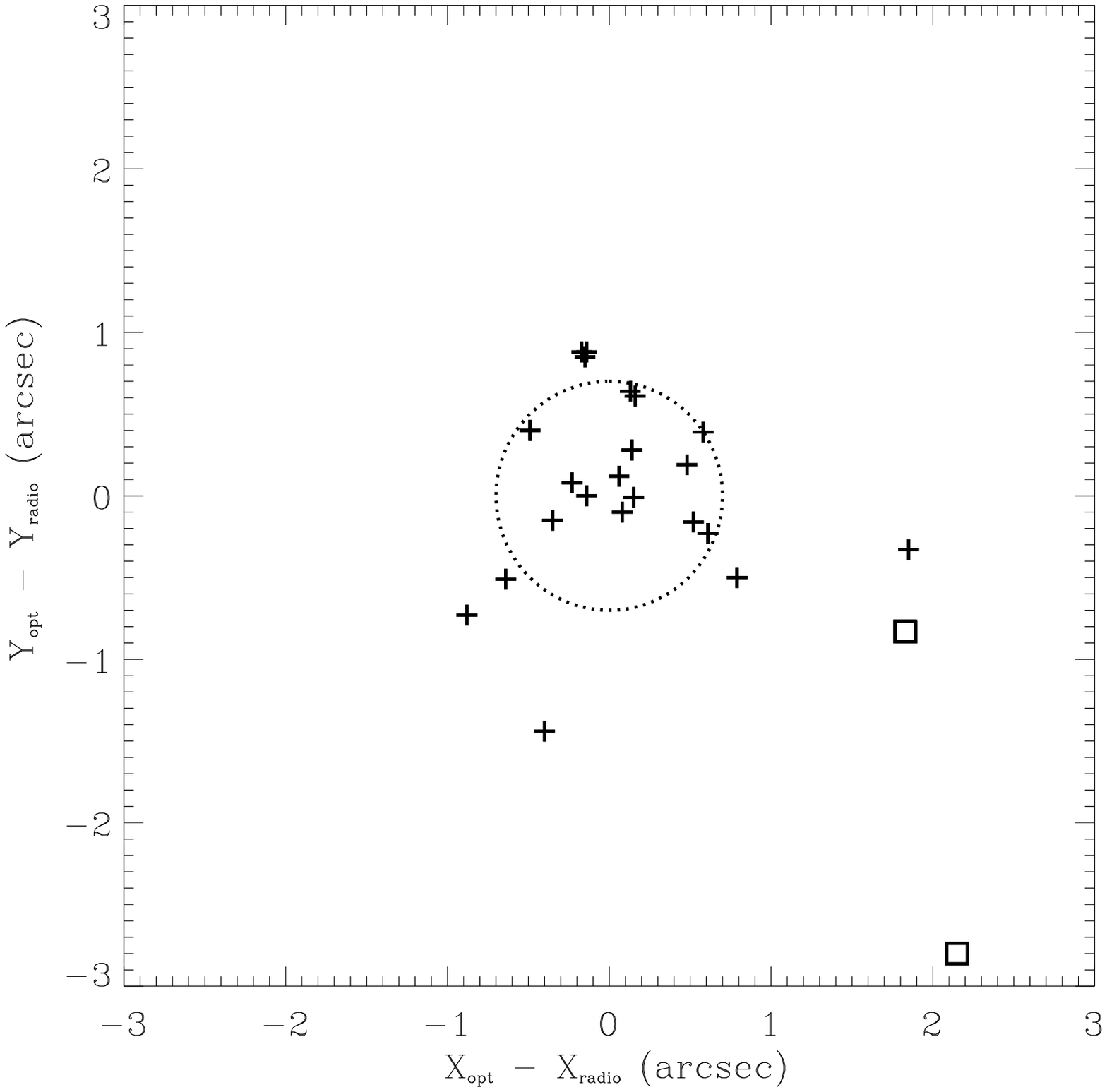,width=8.5cm}
\psfig{figure=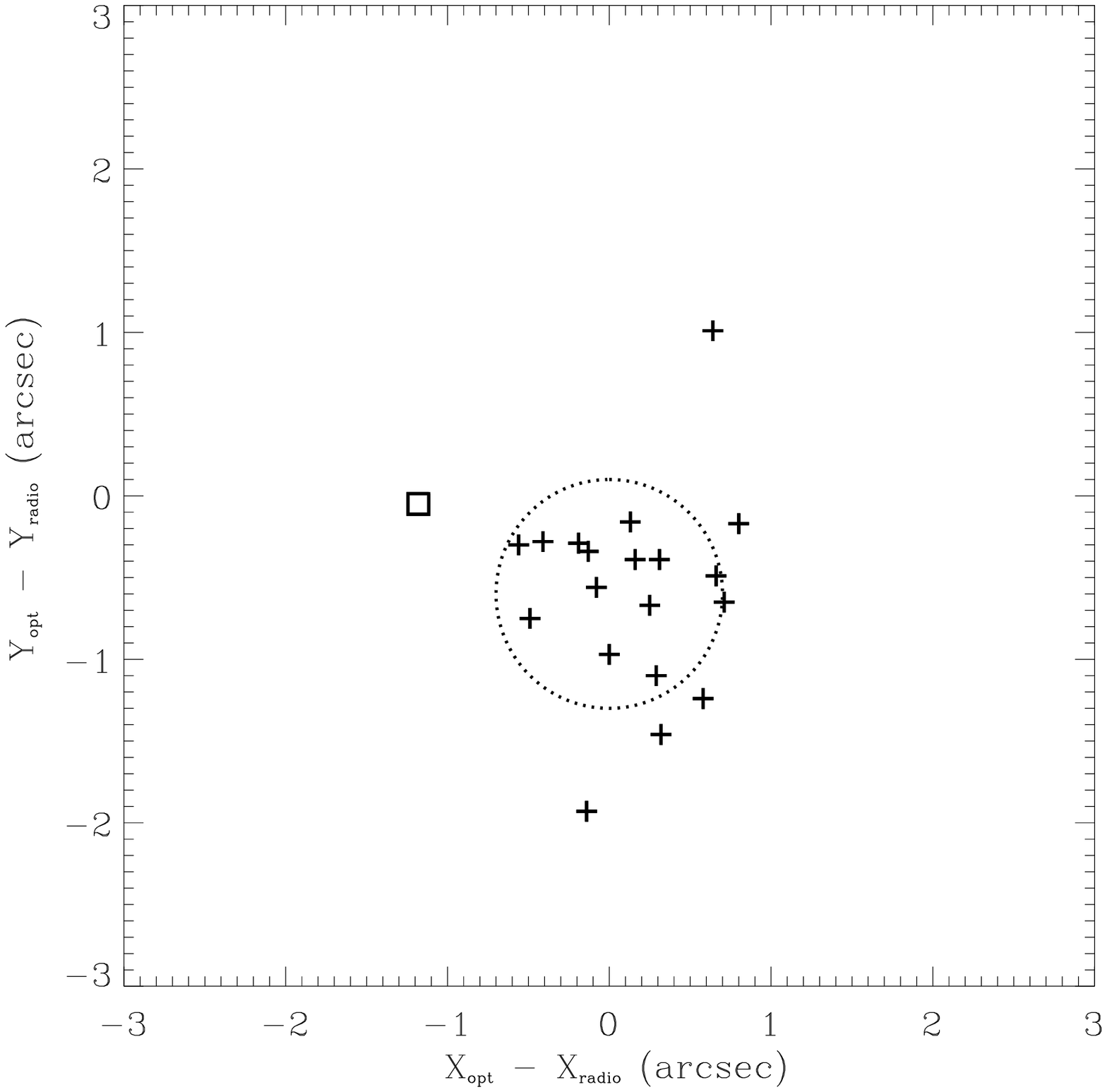,width=8.5cm}}
\caption{\label{pos} The radio-optical position offsets in 
arcseconds for the candidate identifications in the mini-survey region (left)
and in the region $4^h00^m < \alpha < 8^h30^m$ (right). 
The squares indicate candidate
identifications for which there are other objects closer to the radio position.
Note that there is a median offset in declination of $0.6''$ for the 
sources in the region $4^h00^m < \alpha < 8^h30^m$, which has been removed
before determining the dimensionless position errors. The dotted circles 
indicate the uncertainties of the optical position.}
\end{figure*}

An optical object is considered to be a candidate identification if it
is located within an angular distance $3.5''$ from the radio position. 
Only six sources
did not have a candidate identification in the optical or near-infrared.
The radio-optical offsets are shown in figure \ref{pos} for the sources
in the mini-survey region and in the region $4^h00^m < \alpha < 8^h30^m$.
For the sources in the region $4^h00^m < \alpha < 8^h30^m$ and
$58^\circ< \delta < 75^\circ$, a median position offset of $0.6''$ in
declination was found between the optical and the radio. This was
 removed before determining the dimensionless position uncertainties.
The cause of this offset is not clear.  
The median position offsets in $\alpha$ for the region
$4^h00^m < \alpha < 8^h30^m$, and in both $\alpha$ and $\delta$
for the sources in the mini-survey region were found to be smaller than
$0.05''$ and therefore not taken into account.

The background density of objects, $\rho_{bg}$, on the CCD frames, depends
mainly on the galactic latitude and magnitude limit of the observation, and was
in each case estimated by counting the number of objects on the CCD frames 
for each source. The background densities range from 5 to $\sim40$ objects
per square arcminute. The likelihood method assumes that a radio
source and its optical identification always coincide in position, and that
the true identification is always closer to the radio source than the first
contaminating object. Given the normalized position difference $r$ for a
certain radio optical pair, the likelihood ratio is defined by (de Ruiter et
al. 1977);

\begin{equation}
LR(r) = \frac{dp(r|id)}{dp(r|c)} = \frac{r \  e^{\frac{-r^2}{2}}}{2 \lambda \
  r \ e^{-\lambda r^2}} = \frac{1}{2\lambda} e^{\frac{r^2(2\lambda-1)}{2}}
\end{equation}
where $\lambda = \pi \sigma_{\alpha opt} \sigma_{\delta opt} \rho_{bg} =
1.54 \rho_{bg}$. This is the ratio of the probability that a given object
found between $r$ and $r+dr$ is the correct identification $p(r | id)$, divided
by the probability that it is a contaminating object $p(r | c)$.
We have used a likelihood ratio cut off of 2.0, which means that a source
is considered to be identified if the probability that the given object
is the correct identification is twice the probability that it is a 
contaminating background source. A likelihood ratio of 2.0 corresponds to 
an optical to radio offset of $2.9''$ for a background density of 10 objects
per square arcminute. Note that all the candidate identifications
fall within this limit, except for B1647+6225 which has two candidate
counterparts,
of which the closest one is  selected as a possible identification. 

Using a likelihood ratio cutoff of 2.0, we find an identification fraction
$\theta$ of 87\%. This identification fraction can be related to the two 
{\it a posteriori} probabilities that the object found at an angular 
distance $r$ from
the radio source position is a genuine identification, $p(id|r)$, or a 
confusing object, $p(c|r)$, using Bayes' theorem (eg. R\"ottgering et al. 1995).

\begin{equation}
p(id|r)=\frac{\theta \ LR(r)}{\theta \ LR(r) + 1 - \theta}   \ \ \ \ \ 
p(c|r)=\frac{1}{\theta \ LR(r) + 1 - \theta}
\end{equation}

The completeness, $C$, of the identifications, which is the number of accepted identifications
over the number of correct id's, and the reliability, $R$, which is the 
fraction of identified sources
for which the identification is correct, are given by:

\begin{equation}
C=1-\biggl( \sum_{LR(r_i)<L} p(id|r)  \biggr) / N_{id} 
\end{equation}
\begin{equation}
R=1-\biggl( \sum_{LR(r_i)>L} p(c|r)  \biggr) / N_{id}
\end{equation}

Sources in our sample which remain unidentified have no optical and/or
near-infrared counterpart within $3.5''$ of the radio position. This
corresponds to a completeness of $C>99.9\%$. The reliability is determined to
be 97.6\%, which implies that it is probable that one source will have an
erroneous identification.

\subsection{Parameters of the Optical Counterparts}

The magnitudes enclosed within a circular aperture were measured for all the
candidate identifications within various radii from the peak brightness of the
object. These radii are between $3.3''$ and $24''$, depending on the angular
size of the optical identification. Furthermore, for the sources with 
$m_R<21.5$,
the FWHM of the object was measured and compared to that of stars in the 
frame, in order determine if it is an unresolved
or an extended source. A $3\sigma$ detection limit in a $3''$ aperture is
taken as an upper limit for sources without a candidate identification. The
measured magnitudes were then corrected for the galactic foreground extinction
using data from Burnstein and Heiles (1984) derived from HI column densities,
assuming an extinction curve proportional to $\lambda^{-1}$. This might be 
too large a correction in the near-infrared, however extinction correction are
small at those wavelengths anyway. 
These extinction data have been taken from the NASA/IPAC extragalactic 
database (NED).

\section{Results and Discussion}

The results for the individual sources are presented in table \ref{results}.
Column 1 gives the source name; columns 2 the optical positions (J2000)
for the candidate identifications; columns 3 and 4 the radio-optical
position offsets in right ascension and declination; column 5 the dimensionless position error; column 6 gives the morphological classification of the 
identification 
(F=Faint $R>21.5$, S=Star-like, E=Extended), column 7 the galactic 
foreground extinction
in $B$ band, column 8 the aperture diameter used for measuring the magnitudes
and column 9, 10, 11 and 12 give the $R$, $I$, $J$, and $K$ band magnitudes respectively.

The optical and near-infrared images are presented in the appendix, with notes
on individual sources. All the $R$ band
and WHIRCAM $K$ band, and the $I$ band images with detections are shown.
The near-infrared maps taken with ARNICA are shown by Villani and di Serego 
Alighieri (1998). The
$R$ and $I$ band images show a region of $1'\times1'$, and the $K$ band images
a region of $30''\times30''$ around the radio position. To reduce the effect
of digitization of the contours, the data was convolved with a Gaussian with a
FWHM of 3 pixels. The contours are at 3, 6, 12, 24, 48, etc. times the sky
noise for the bright sources, and at 3, 4, 6, 12, 24, 48 etc. times the 
sky noise for the faintest sources. The level of the faintest contour 
 in mag/arcsec$^2$ is given at the top-right corner of each image.

\subsection{Magnitude and Colour Distributions}

\begin{figure}
\psfig{figure=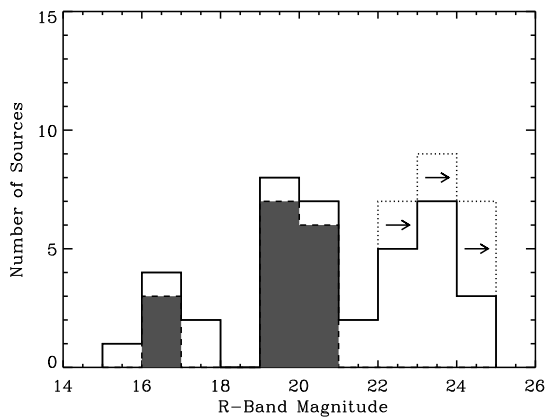,width=8cm}
\caption{\label{magdis} The $R$ band magnitude distribution 
of the sources in the sample. The shaded areas indicate 
identifications with star-like objects, which are assumed to be quasars.
The arrows indicate lower limits.}
\end{figure}

Figure \ref{magdis} shows the $R$ band magnitude distribution of the
sources in the sample. If a radio source has more than one candidate
identification, the one closest to the radio position was taken as the
ID. The shaded areas indicate identifications with star-like objects,
which are assumed to be quasars.  In our sample there are 6 bright
galaxies (13$\pm5$\%), 20 bright quasars (43$\pm9$\%), and 21 faint objects (44$\pm10$\%) for which morphological classification is uncertain. 
This was compared with the radio bright sample of Stanghellini et al. (1998)
containing 33 sources, which is complete to a flux density of 1 Jy at 5 GHz.
For this radio bright sample, the percentages are $36\pm10$\% for the bright
galaxies, $42\pm11$\% for the quasars, and $21\pm8$\% for the faint objects
with $m_R>21.5$. While the percentages of bright quasars in both samples 
are the same within the errors, the ratio of bright to faint
galaxies is lower in our faint sample than in the radio bright 
sample at a $2\sigma$ level. Assuming that the faint optical objects are 
galaxies, this may indicate that the redshift distribution of GPS galaxies in 
our sample is more biased towards higher redshift.

\begin{figure}
\centerline{
\psfig{figure=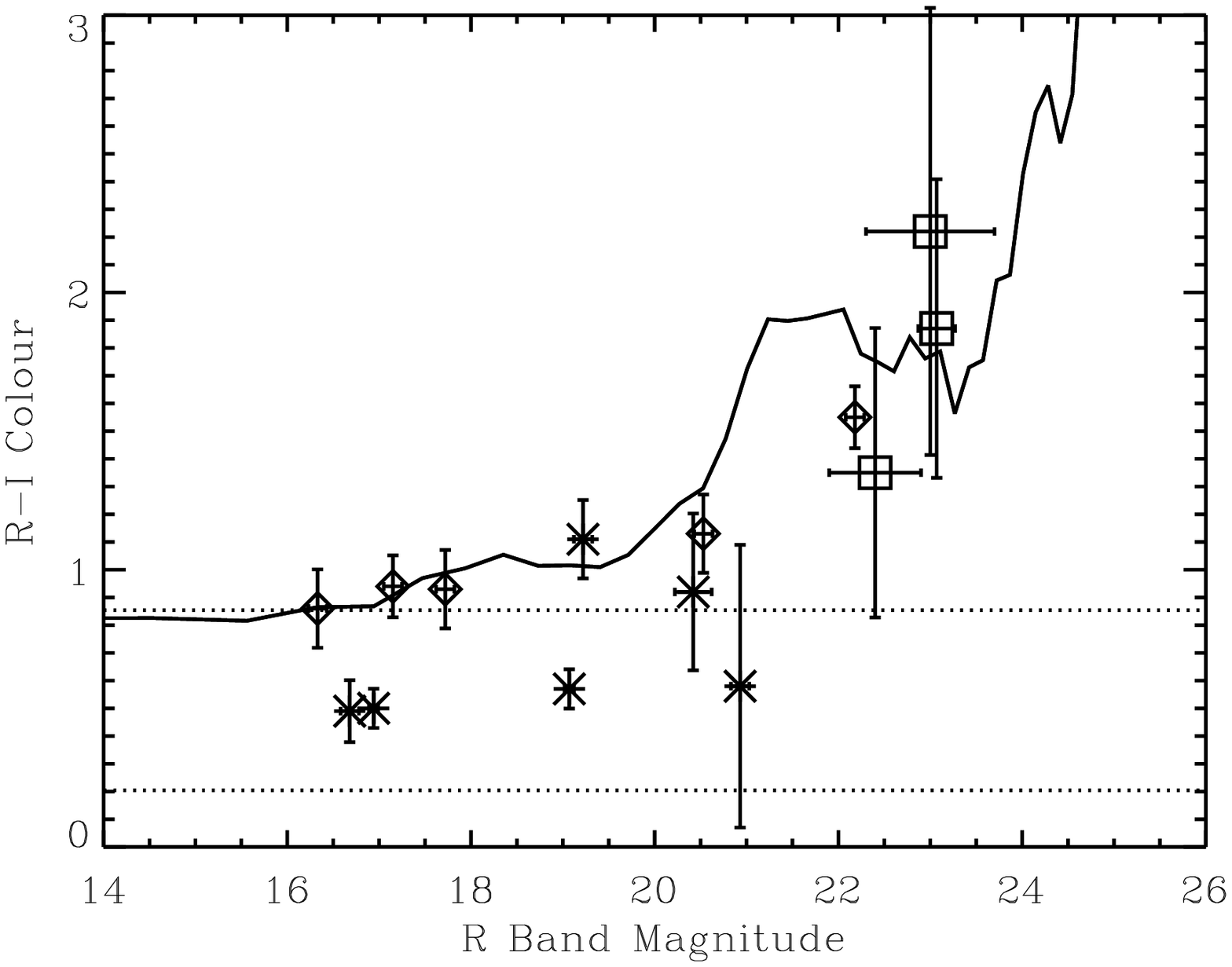,width=9cm}}
\centerline{
\psfig{figure=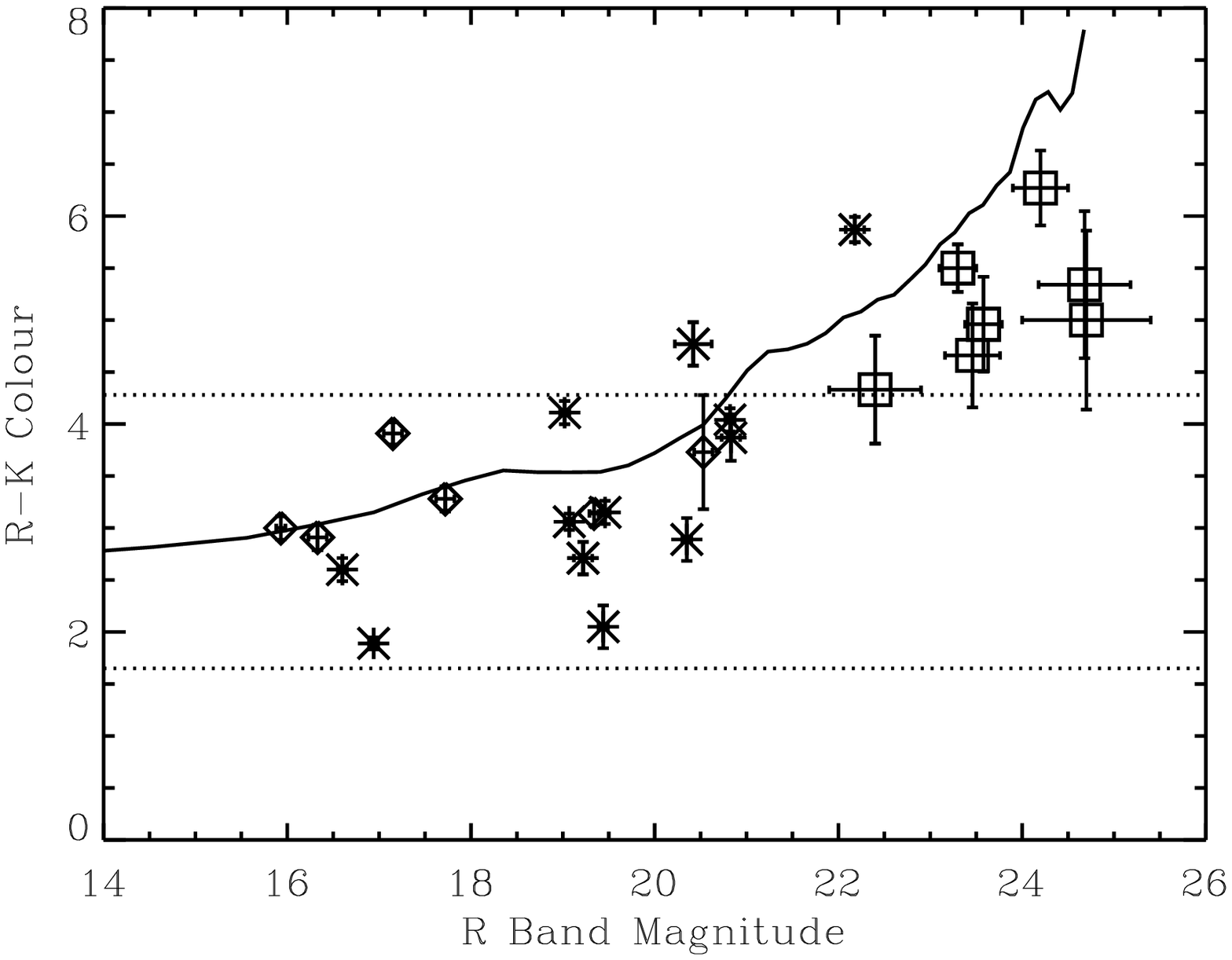,width=9cm,clip=}}
\caption{\label{colours} $R-I$ (top) and $R-K$ (bottom) colours 
as function of $R$ band magnitude.
The diamonds represent the extended objects, the
crosses the stellar objects, and the squares the faint $m_R>21.5$
identifications. The dotted lines show the  upper and lower limits on colour 
for the expected range of spectral energy distributions for quasars.
The solid lines indicate the expected colours for 
passively evolving ellipticals which are on the $R$ band Hubble diagram 
found for radio bright GPS galaxies (Snellen et al. 1996a).}
\end{figure}

Figure \ref{colours} shows the available $R-I$ and $R-K$ colours as
function of $R$ band magnitude. The diamond symbols represent the
extended objects, the crosses the star-like objects, and the squares
the faint $m_R>21.5$ identifications. The dotted lines show the upper
and lower limits on colour for a spectral energy distribution of
$F_{\nu} \propto \nu^{-\alpha_{opt}}$ with $0.0>\alpha_{opt}>-2.0$,
which is the expected range for $\alpha_{opt}$ of radio loud quasars
(eg. Baker and Hunstead 1995).
All the colours of the stellar objects are within the expected range for 
quasars, except for two $R-I$ colours and one $R-K$ colour. These are slightly redder than expected which may be due to the contribution of emission
lines in the $I$ or $K$ band.
 
The solid lines represent the $R-I$ and
$R-K$ colours expected for passively evolving ellipticals, assuming
that the galaxy magnitudes lie on the $R$ band Hubble diagram for GPS
sources defined in Snellen et al. (1996a). 
The $R-I$ colours of the extended and faint identifications are
consistent with this, but four from the six faint objects with
measured $R-K$ colours are bluer ($> 2 \sigma$) than expected. 
These could still be
passively evolving ellipticals with high formation redshifts as 
long as they are intrinsically fainter than expected on the basis of 
the $R$ band Hubble relation.
Indeed the Hubble diagram is poorly determined for galaxies with $R>23.0$. 
The discrepancy of 1.5 magnitudes in $R-K$ colour for a $m_R=24.0$ galaxy
could be explained if it is not at a redshift of 1.3, but at a
redshift of 1.0. This is well within the uncertainties of the Hubble
diagram at such a high redshift. 

\subsection{The Optical Morphologies and Environments}

\begin{figure*}
\hbox{
\psfig{figure=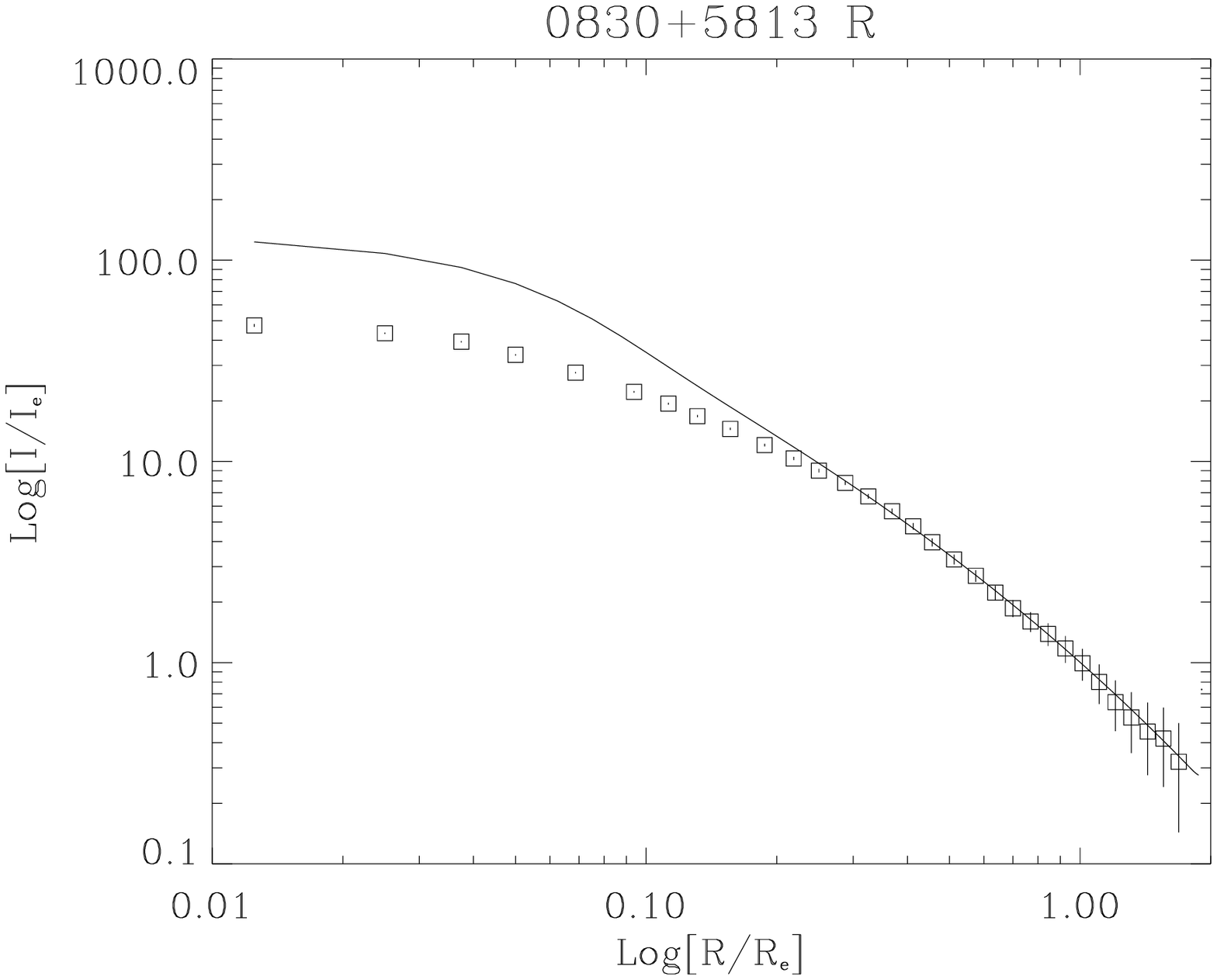,width=8.7cm}
\psfig{figure=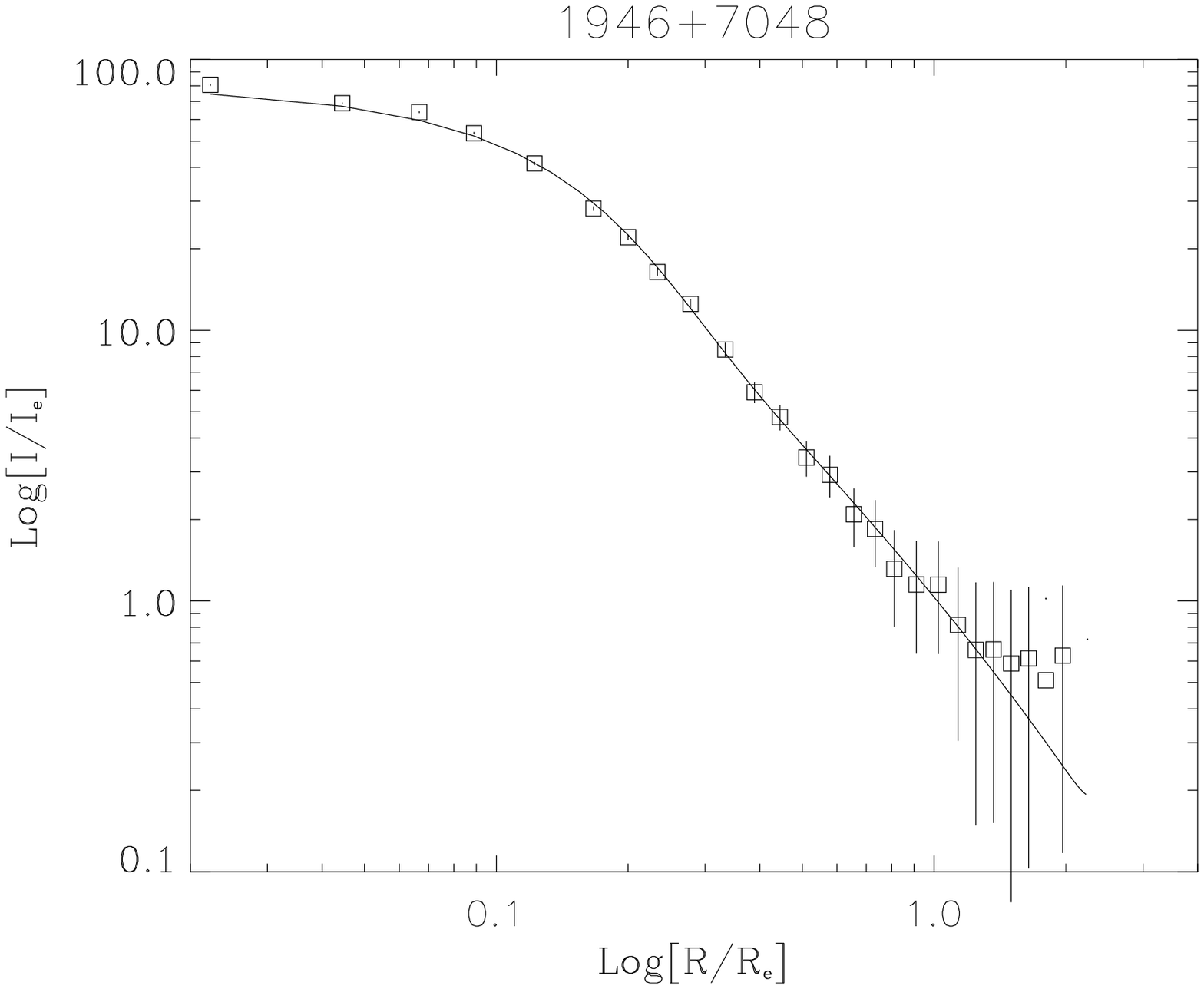,width=8.7cm}}
\caption{\label{prof} The angular profiles of B0830+5813 and
B1946+7048, fitted with an $R^{1/4}$ law. Note that 
the profile of B0830+5813
clearly flattens in the inner 10 pixels indicating that  
light is missing in the centre of the galaxy, which may be due to the 
presence of dust.}
\end{figure*}

It has been claimed by Stanghellini et al. (1993) that 75\% (8 out of 11) of
the GPS galaxies in the radio bright sample with $m_R<20.5$ have distorted
isophotes, double nuclei or close neighbours. There are four galaxies in our
sample which have $m_R<20.5$ and can be used to check this claim. The angular 
morphologies of B0830+5813, B1622+6630, B1819+6707 and B1946+7048 have 
ellipticities of 0.173, 0.026, 0.070, and 0.031, with position angles of 
$29^\circ $, $61^\circ $, $53^\circ $, and $51^\circ$ respectively.
Contour plots
of the two brightest galaxies, B0830+5813 and B1946+7048, seem to indicate that these
are ``relaxed'' systems (as indicated by the regular symmetric contours). 

To check this  an $r^{1/4}$ law was fitted to the galaxy profiles 
(figure \ref{prof}). The profile of B1946+7048 is well fitted by an $r^{1/4}$ law,
with an effective radius of $R_{eff} = 13''$, but the profile of B0830+5813
clearly flattens in the inner 10 pixels. In this case an $r^{1/4}$
law was fitted to the outer part of the profile ($R_{eff} = 14''$). 
This indicates that about half of the expected 
light is missing in the centre of the galaxy, possibly due to the 
presence of dust. Unfortunately the $K$ band data on this
galaxy is of too poor a quality to measure the colour gradient along 
the profile. Assuming that the light profile without dust followed the 
$R^{1/4}$ law, then an estimate of the dust mass $M_{dust}$ can be made.
Following van Dokkum and Franx (1995), the dust mass corresponds to 
\begin{equation}
M_{dust} = \Sigma \bar{A_V} \Gamma _V^{-1}
\end{equation}
where $\Sigma$ is the area covered by the dust, $\bar{A_V}$ is the average
extinction in this area, and $\Gamma _V \approx 6\times10^{-6}$ mag kpc$^2$
$M_{\sun} $ visual mass absorption coefficient.
The average extinction is estimated to be 0.8 mag within the radius 
$R_{dust}=1.4''$. The redshift of the galaxy is z=0.093 
(Snellen 1997, Snellen et al. in prep), hence this corresponds to an area 
of 30 kpc$^2$ ($H_0 = 50$ km/sec/Mpc, $q_0 = 0.5$).
This gives a dust mass estimate of
$M_{dust} \approx 10^{6.5}$ $M_{\sun}$. This is comparable to dust-masses
estimated in 3C galaxies (de Koff et al., 1998).

The exposure times on  B1622+6630 and B1819+6707 were not long enough
to construct  reliable profiles such as for B0830+5813 and B1946+7048,
but there are indications from the noisy profiles that they are also
relaxed, although a
nuclear point source might well be present in B1622+6630.

\begin{figure*} 
\centerline{
\psfig{figure=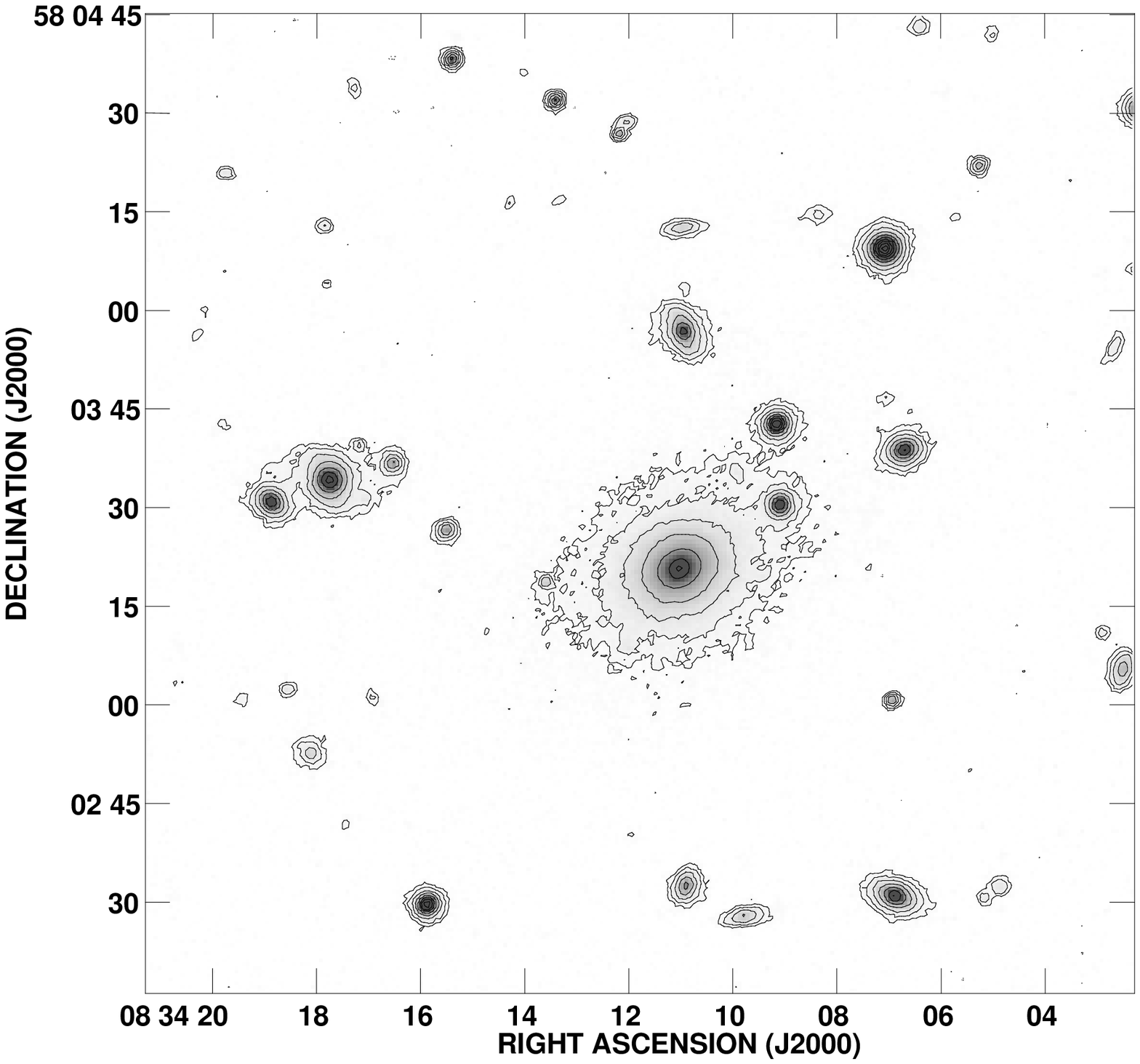,width=14cm}}
\caption{\label{cluster} The GPS galaxy B0830+5813, which is clearly
located in a cluster environment}
\end{figure*}

Of the four brightest galaxies, two seem to have close neighbours but
none of them show evidence for an ongoing galaxy interaction. There
are two objects located at a distance of $\sim30''$ from the core of
B0830+5813, clearly within the envelope of the galaxy. Although these
objects may well not be associated physically with the main galaxy and
the result of a projection effect. B0830+5813 seems to be located in a
cluster environment (fig. \ref{cluster}). There are 13 galaxies on the
CCD image with magnitudes between the magnitude of the third brightest
galaxy and 2 magnitudes fainter. This corresponds to an Abell richness
class (Abell, 1958) of $2-3$, assuming a modified Hubble profile
(Sarazin, 1986) for the projected surface density of galaxies with a
core radius of 500 Kpc. The galaxies B1622+6630 and B1819+6707 seem to
have close neighbours at a distance of respectively 2.0$''$ and
1.5$''$, but B1946+7048 has no neighbouring galaxies.

\subsection{The Optical Identifications and their Radio Spectra}

\begin{figure} 
\centerline{
\psfig{figure=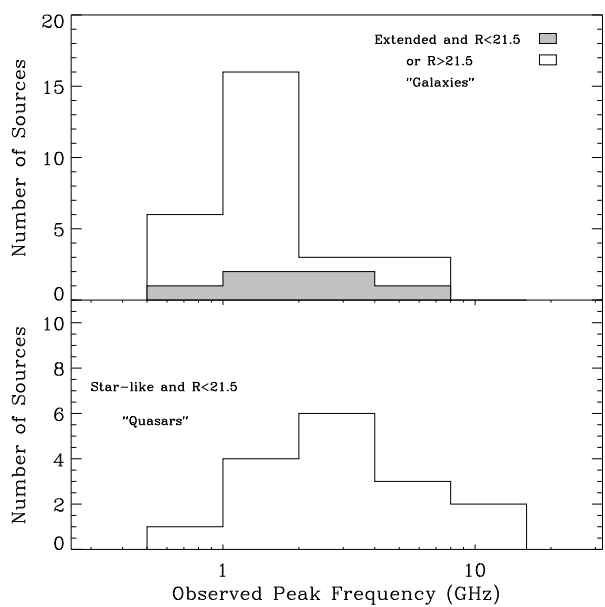,width=9cm}}
\caption{\label{PVID} The radio peak frequency distribution for 
the faint or extended objects, assumed to be galaxies, and the bright
stellar objects, assumed to be quasars. The two distributions are different
at a 99\% confidence level (KS-test). }
\end{figure}

\begin{figure}
\centerline{
\psfig{figure=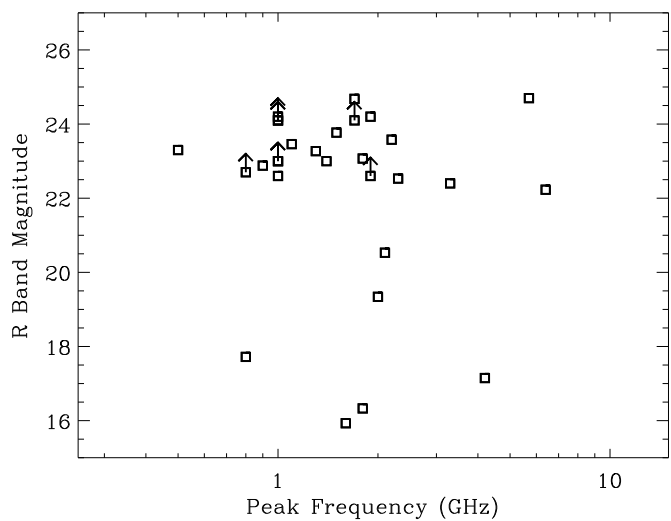,width=10cm}}
\caption{\label{RPV} The observed radio peak frequency as function of 
$R$ magnitude for the faint and/or extended objects, which are assumed to
be the GPS galaxies. There is no correlation found.}
\end{figure}

If the orientation unification scheme for extragalactic radio sources
is correct, then it is expected that the redshift distributions of
radio galaxies and quasars are more or less the same. If differences occur
then these must be due to luminosity effects, for example, a change of quasar
opening angle with radio power (eg. Urry and Padovani, 1995). 
The observed redshift distributions in radio-bright samples of GPS sources identified 
with galaxies and quasars are found to be very different, with the GPS quasars
located at high redshifts ($2<z<4$, O'Dea et al, 1991).  However,
two selection effects influence this result. Firstly, the GPS galaxies
are biased towards low redshift because spectroscopy is easier on low
redshift galaxies. Secondly, the GPS quasars may be biased towards high
redshift because their intrinsic peak frequencies seem to be higher (eg. de Vries et al 1997, Stanghellini et al 1996).
If that is true, low redshift GPS quasars may fall outside the bright
 samples because their observed peak frequencies are too
high.  We have found evidence that the redshift distribution of GPS
galaxies in the faint sample is indeed different from that of the GPS
quasars (Snellen 1997).  The faint sample is more suitable for this,
because sources are selected on the optically thick part of their radio
spectrum, and moving sources within the sample to lower redshift,
thereby shifting the observed peak frequency to lower frequencies,
will not make them disappear from the sample. 
The difference in redshift distributions is
unlikely to be caused by a radio power effect, because it occurs both in the
faint and bright samples. This seems to indicate
that GPS galaxies and quasars are not unified by orientation and form
two different classes of objects which happen to have the same radio
spectral properties.

As mentioned above, it has been found (Stanghellini et al., 1996, 
de Vries etal. 1997) that the
intrinsic peak frequency distributions of GPS quasars and galaxies are
different. However, the selection bias discussed above may be
responsible for this. We conclude that it would be better to compare the 
{\it observed}
peak frequency distributions of the galaxies and quasars. Although
radio selection effects are severe in the observed frame, they are
independent of optical host. If all optical
counterparts of the faint GPS sources that are classified as
``extended'' or ``faint'' can be assumed to be galaxies, and all unresolved
star-like identifications quasars, figure \ref{PVID} shows that there is a 
clear difference in the peak frequency distribution for these two
groups. The hypothesis that the distributions of the quasars and
galaxies are drawn from the same parent distribution can be rejected
at $>99\%$ confidence level using the Kolmogorov-Smirnov test.

De Vries et al. (1997) have found an apparent absence of sources at
high redshift with low peak frequencies, which they interpreted as a
correlation between redshift or radio luminosity with peak frequency.
We believe this is not a dependence of redshift, but a dependence on
optical host because the quasars tend to be at higher redshift than the 
galaxies. To check this we investigated whether the observed peak
frequency is a function of redshift for the GPS galaxies only. The
redshifts are not known for the majority of the GPS galaxies, but the
$R$ band magnitude of GPS galaxies is a good indicator of the
redshift, at least for the radio bright samples (Snellen et
al. 1996a).  Plotting the $R$ band magnitude of the `candidate'
galaxies versus their radio peak frequency however, shows no
correlation between these two quantities (fig. \ref{RPV}).

\section{Conclusions}

A sample of 47 faint GPS sources selected from the Westerbork Northern
Sky Survey has been imaged in the optical and near-infrared, resulting
in an identification fraction of 87\% with a reliability of 97.6\% and
a completeness of $>99.9\%$. The optical to near infrared colours of
the faint and the extended optical identifications are found to be in
agreement with passively evolving elliptical models, assuming that their
redshifts can be determined from the $R$ band Hubble diagram as defined
for radio bright GPS galaxies (Snellen et al. 1996a), and taking into
account the large uncertainty in the Hubble diagram at high redshift.

The GPS sources optically identified with bright stellar objects,
presumably quasars, have in the observed frame, on average, higher
radio peak frequencies than the GPS sources identified with faint and
extended objects, presumably galaxies.  This indicates that there is a
difference between the true peak frequency distributions of GPS
galaxies and quasars, because selection effects are the same for both
type of optical counterparts. Such a difference in 
intrinsic peak frequency distribution may cause a selection bias,
resulting in a different redshift distribution for GPS galaxies and
quasars.  However, it has been shown by this faint sample (Snellen
1997) that this is not the case, and that {\it both} the redshift
distribution and intrinsic peak frequency distribution for GPS galaxies and
quasars are different.

This seems to indicate that GPS galaxies and quasars are not unified
by orientation and form two different classes of objects which happen
to have the same radio spectral properties.

\section*{acknowledgements}
We thank the {\it Comite Cientifico International} (CCI) of the IAC
for the allocation of observing time. 
The Jacobus Kapteyn Telescope, the Isaac Newton Telescope, and the William 
Herschel Telescope are
operated on the island of La Palma by the Isaac Newton Group in the Spanish
Observatorio del Roque de los Muchachos of the Instituto de Astrofisica de 
Canarias. We thank Daniela Villani and Sperello di Serego Alighieri for
providing unpublished data. The NASA/IPAC Extragalactic Database (NED) is 
operated by the Jet Propulsion Laboratory, California Institute of Technology,
 under contract with the National Aeronautics and Space Administration. 
This work was in parts funded
through a NWO programme subsidy and by the European Commission under
contracts SCI*-CT91-0718 (The Most Distant Galaxies) and 
ERBFMRX-CT96-086 (Formation and Evolution of Galaxies), and
ERBFMRX-CT96-0034 (CERES). RGM thanks the Royal Society for support.

{}

\section*{Appendix A}
\noindent {\bf Comments on individual sources}

\noindent {\bf B0400+6042} No optical identification has been found to a limiting magnitude of $R=22.1$ and $K=18.5$.

\noindent {\bf B0436+6152} No optical identification has been found to a limiting magnitude of $R=22.1$ and $K=18.5$.

\noindent {\bf B0441+5757} A very bright star is located only $\sim 15''$ away
to the north-west. There may be a second 
object to the south east very close $\sim 2''$ to the identification. 

\noindent {\bf B0539+6200} An optical identification is found in $R$ band, 
just at the detection limit. However, the object is clearly detected in 
$K$ band.

\noindent {\bf B0552+6017} No optical identification has been found to a limiting magnitude of $R=23.8$.

\noindent {\bf B0759+6557} This object has faint detections in both 
$R$ and $K$ band.

\noindent {\bf B0826+7045} Two possible identifications are found at 0.8 and
1.2 arseconds distance from the radio position.

\noindent {\bf B1600+7131} No optical identification has been found to a limiting magnitude of $R=24.1$ and $I=21.5$.

\noindent {\bf B1620+6406} A very bright star is located only 2 arcminutes
to the south west.

\noindent {\bf B1622+6630} Two objects are found within an angular distance 
of 2$''$. The may be neighbouring galaxies.

\noindent {\bf B1639+6711} No optical identification has been found to a limiting magnitude of $R=24.1$ and $I=21.5$. 

\noindent {\bf B1647+6225} Two objects are found within 3.5 arseconds from
the radio position. The fainest one is probably the real identification.

\noindent {\bf B1655+6446} No optical identification has been found to a limiting magnitude of $R=24.1$ and $I=21.5$.

\noindent {\bf B1807+5959}  A possible identification is found 
at a $3\sigma$ magnitude level.

\noindent {\bf B1807+6742} No optical identification has been found to a limiting magnitude of $R=22.7$ and $I=21.5$.

\noindent {\bf B1819+6707} Two objects are found within an angular distance 
of 1.5$''$. The may be neighbouring galaxies.

\noindent {\bf B1843+6305} This source has no identification to $R>22.6$ 
but is clearly detected in $I$ band.

\noindent {\bf B1942+7214} The optical identification is very faint in $R$ 
but is clearly visible in $I$ and $J$ band.

\noindent {\bf B1954+6146} The optical identification is a Red point source, 
with very faint extended emission in $R$ band.

\noindent {\bf B1958+6158} Two objects are found within 2.0 arseconds from
the radio position.

\begin{table*}
\setlength{\tabcolsep}{1.0mm}
\renewcommand{\arraystretch}{0.95}
\begin{tabular}{|c|rrrrrrrrc|ccccccc|}\hline
Source &\multicolumn{6}{c}{Optical Position (J2000)}& RA & 
Decl&r &ID &$A_B$& Ap. & $R_{ap}$ & $I_{ap}$ & $J_{ap}$ &$\ \ \  K_{ap}\ \ \ $\\
       &\multicolumn{3}{c}{R.A.}&\multicolumn{3}{c}{Decl.}&off. &off.&&&&&&&&    \\
       &h&m&s&$\circ$&$'$&$''$&$''$&$''$&& & &$''$&mag.&mag.&mag.&mag.\\ 
   (1)       & \multicolumn{6}{c}{(2)} & (3)  & (4) &(5)&(6)&(7)&(8)&(9)&(10)&(11)&(12)\\ \hline
B0400+6042&  &   &      &    &   &      &     &     &   &-&$>2.7$&3.0 &$>22.1$        &           &&$>18.5       $\\
B0436+6152&  &   &      &    &   &      &     &     &   &-&2.45&3.0 &$>22.1$        &             &&$>18.5       $\\
B0441+5757&04& 46&  8.82&  58& 02& 48.92&+0.6&+1.0&1.7&S&$>2.3$&5.3&$16.60\pm0.05$ &              &&$14.00\pm0.10$\\
B0513+7129&05& 19& 29.01&  71& 33&  2.84&+0.3&$-$1.1&1.6&S&0.49&5.3&$20.82\pm0.05$ &              &&$16.78\pm0.10$\\
B0531+6121&05& 36& 30.88&  61& 23& 22.31&+0.6&$-$1.2&2.0&E&0.99&7.0&$19.02\pm0.05$ &              &&$14.91\pm0.10$\\
B0535+6743&05& 41& 13.43&  67& 45& 22.36&+0.0&$-$1.0&1.4&F&0.39&5.3&$24.70\pm0.70$ &              &&$19.70\pm0.50$\\
B0537+6444&05& 42& 10.02&  64& 46& 31.99&$-$0.2&$-$0.3&0.5&S&0.55&5.3&$19.46\pm0.05$ &            &  &$16.31\pm0.10$\\
B0538+7131&05& 44& 31.37&  71& 32& 40.60&$-$0.5&$-$0.8&1.3&S&0.41&4.2&$20.83\pm0.10$ &            &  &$16.96\pm0.20$\\
B0539+6200&05& 44& 33.57&  62& 01& 17.86&$-$0.1&$-$1.9&2.7&F&0.73&5.3&$24.20\pm0.30$ &            &  &$17.93\pm0.20$\\
B0543+6523&05& 48& 39.79&  65& 24& 24.43&+0.3&$-$0.7&1.1&S&0.58&5.3&$20.35\pm0.05$ &              &&$17.46\pm0.20$\\
B0544+5847&05& 48& 27.01&  58& 47& 55.51&+0.2&$-$0.4&0.6&S&1.02&5.3&$19.44\pm0.05$ &              &&$17.39\pm0.20$\\
B0552+6017&  &   &      &    &   &      &     &     &   &-&0.57&3.0&$>23.8$        &              &&$>18.9       $\\
B0557+5717&06& 01& 49.46&  57& 17& 20.82&$-$0.4&$-$0.3&0.7&F&0.97&4.2&$23.46\pm0.30$ &            &  &$18.80\pm0.40$\\
B0601+5753&06& 05& 42.29&  57& 53& 14.87&+0.3&$-$1.5&2.2&S&0.59&5.3&$19.08\pm0.05$ &              &&\\
B0748+6343&07& 53&  0.76&  63& 35& 45.61&+0.7&$-$0.5&1.2&S&0.12&5.3&$21.20\pm0.10$ &              &&\\
B0752+6355&07& 56& 54.66&  63& 47& 58.71&+0.1&$-$0.2&0.3&F&0.11&3.5&$22.23\pm0.30$ &              &&\\
B0755+6354&07& 59& 52.86&  63& 46& 13.03&$-$0.1&$-$0.3&0.5&S&0.11&4.2&$19.08\pm0.10$ &            &  &\\
B0756+6647&08& 01& 36.51&  66& 39&  9.04&+0.7&$-$0.7&1.4&S&0.12&5.3&$20.44\pm0.05$ &              &&\\
B0758+5929&08& 02& 24.59&  59& 21& 34.00&$-$0.1&$-$0.6&0.9&S&0.18&5.3&$19.34\pm0.05$ &            &  &$16.20\pm0.10$\\
B0759+6557&08& 03& 55.52&  65& 49& 18.46&+0.3&$-$0.4&0.7&F&0.18&3.5&$24.68\pm0.50$ &              &&$19.34\pm0.50$\\
B0826+7045&08& 31& 57.51&  70& 35& 35.96&$-$1.2&$-$0.1&1.7&S&0.08&4.2&$17.20\pm0.05$ &            & &\\
          &08& 31& 57.91&  70& 35& 35.84&+0.8&$-$0.2&1.3&S&0.08&2.1&$19.65\pm0.15$ &              &&\\
B0830+5813&08& 34& 11.03&  58& 03& 20.75&$-$0.6&$-$0.3&1.0&E&0.21&21.1&$15.93\pm0.05$&            &  &$12.93\pm0.10$\\ 
B1525+6801&15& 25& 45.95&  67& 51& 24.39&$-$0.2&+0.9&1.3&F&0.05&4.7&$23.07\pm0.20$ &$21.20\pm0.50$&&              \\
B1538+5920&15& 39& 29.45&  59& 11&  0.45&+0.2&$-$0.0&0.3&S&0.04&4.8&$20.93\pm0.10$ &$20.35\pm0.50$&&              \\
B1550+5815&15& 51& 58.13&  58& 06& 44.92&+0.2&+0.6&0.9&S&-&5.3&$16.68\pm0.10$ &$16.19\pm0.05$&    &             \\
B1551+6822&15& 52&  4.22&  68& 13& 46.81&$-$0.9&$-$0.7&1.6&F&0.05&3.6&$23.77\pm0.30$ &$>21.5$     &  &              \\
B1557+6220&15& 57& 52.73&  62& 11& 37.27&$-$0.5&+0.4&0.9&F&0.00&4.8&$22.53\pm0.30$ &$>21.5$       &&              \\
B1600+7131&  &   &      &    &   &      &     &     &   &-&0.08&3.0&$>24.1$        &$>21.5$       &&              \\ 
B1620+6406&16& 21& 15.27&  63& 59& 13.63&+0.1&+0.6&0.9&F&0.03&3.3&$23.58\pm0.20$ &$>21.5$       &&$18.62\pm0.41$\\
B1622+6630&16& 23&  4.51&  66& 24&  0.99&+0.5&+0.2&0.8&E&0.07&7.7&$17.15\pm0.10$ &$16.21\pm0.05$&&$13.24\pm0.04$\\
B1639+6711&  &   &      &    &   &      &     &     &   &-&0.12&3.0&$>24.1$        &$>21.5$       &&              \\ 
B1642+6701&16& 42& 21.75&  66& 55& 48.89&$-$0.6&$-$0.5&1.1&S&0.09&4.4&$17.03\pm0.05$ &$16.44\pm0.05$ & &            \\
B1647+6225&16& 48&  3.85&  62& 20& 36.36&+2.2&$-$2.8&5.1&S&0.03&3.9&$20.41\pm0.10$ &$>21.5$       &&$>17.92$\\
         &16& 48&  3.56&  62& 20& 39.44&+0.1&+0.3&0.5&F&0.03&3.3&$22.88\pm0.30$ &$>21.5$       &  &            \\
B1655+6446&  &   &      &    &   &      &     &     &   &-&0.05&3.0&$>24.1$        &$>21.5$       &&              \\
B1657+5826&16& 58&  5.05&  58& 22&  2.19&+0.1&$-$0.1&0.2&F&0.00&3.3&$23.30\pm0.20$ &$>21.5$       &$19.40\pm0.15$&$17.80\pm0.11$\\
B1746+6921&17& 46& 30.04&  69& 20& 35.74&+0.6&+0.4&1.0&S&0.14&3.9&$19.22\pm0.10$ &$18.11\pm0.10$&&$16.51\pm0.12$\\
B1807+5959&18& 07& 55.41&  59& 59& 59.85&$-$0.2&+0.1&0.3&F&0.14&3.6&$22.60\pm0.50$ &$>21.5$       &&              \\
B1807+6742&18& 07& 14.17&  67& 42& 52.84&$-$0.4&$-$1.4&2.1&-&0.19&3.0&$>22.7$        &$>21.4$     &  &              \\
B1808+6813&18& 08& 12.25&  68& 14& 12.08&+1.9&$-$0.3&2.7&F&0.19&3.3&$23.27\pm0.20$ &$>21.4$       &$20.11\pm0.28$&$>18.85$      \\
B1819+6707&18& 19& 44.38&  67& 08& 47.02&+0.5&$-$0.2&0.8&E&0.18&6.6&$17.72\pm0.10$ &$16.79\pm0.10$&$15.85\pm0.03$&$14.44\pm0.07$\\
B1841+6715&18& 41&  3.86&  67& 18& 50.05&+0.1&+0.1&0.2&E&0.30&4.4&$20.53\pm0.10$ &$19.40\pm0.10$& &$16.80\pm0.54$\\
B1843+6305&18& 43& 30.50&  63& 08& 50.80&+0.6&$-$0.2&0.9&-&0.28&3.6&$>22.6$        &$20.56\pm0.30$&&              \\
B1942+7214&19& 41& 27.04&  72& 21& 41.88&+0.8&$-$0.5&1.3&F&0.48&3.3&$23.00\pm0.70$ &$20.78\pm0.40$&$20.36\pm0.17$&              \\
B1945+6024&19& 46& 12.95&  60& 31& 39.97&$-$0.2&+0.9&1.3&S&0.33&3.3&$20.42\pm0.20$ &$19.50\pm0.20$&$17.49\pm0.05$&$15.65\pm0.06$\\
B1946+7048&19& 45& 53.32&  70& 55& 48.86&$-$0.4&$-$0.2&0.6&E&0.53&24.0&$16.33\pm0.10$&$15.47\pm0.10$&$14.23\pm0.24$&$13.42\pm0.30$\\
B1954+6146&19& 54& 55.87&  61& 53& 59.39&$-$0.1&+0.9&1.3&S&0.37&3.3&$22.18\pm0.10$&$20.63\pm0.05$&$18.11\pm0.05$&$16.31\pm0.07$\\
B1958+6158&19& 59& 30.30&  62& 06& 43.19&+1.9&$-$0.8&2.9&S&0.40&3.3&$19.07\pm0.05$ &$18.50\pm0.05$&$17.05\pm0.05$&$16.01\pm0.06$\\
         &19& 59& 30.02&  62& 06& 44.96&$-$0.1&+0.0&0.1&F&0.40&3.0&$22.40\pm0.50 $&$21.05\pm0.15$&$19.57\pm0.18$&$18.07\pm0.15$\\\hline
\end{tabular}
\caption{\label{results} The optical parameters of the candidate identifications.}
\end{table*}

\clearpage

\begin{figure*}

\vspace{21cm}
\caption{{\bf THIS FIGURE IS AVAILABLE AT: 
http://www.ast.cam.ac.uk/~snellen}.
The optical and near-infrared images. Note that the angular scale is 
different for the $K$ band images.}
\end{figure*}

\end{document}